\documentclass[amsmath, twocolumn, showpacs, aps, prl, superscriptaddress]{revtex4-1}
\usepackage{bm}
\usepackage{graphics}
\usepackage{graphicx}
\usepackage{color}
\usepackage{amssymb}
\usepackage{color}

\newcommand{\bsigma}{{\boldsymbol \sigma}}
\newcommand{\bnabla}{{\boldsymbol \nabla}}

\newcommand{\bepsilon}{{\boldsymbol \epsilon}}

\begin{document}

\title{Chiral Electromagnetic Waves in Weyl Semimetals}
\author{Alexander A. Zyuzin}
\affiliation{
Department of Physics, University of Basel, Klingelbergstrasse 82, CH-4056 Basel, Switzerland
}
\affiliation{
A.F. Ioffe Physico-Technical Institute of Russian Academy of Sciences, 194021 St. Petersburg, Russia
}
\author{Vladimir A. Zyuzin}
\affiliation{
Department of Physics, University of Florida, Gainesville, FL 32611-8440, USA
}



\begin{abstract}
We show that Weyl semimetals with broken time-reversal symmetry can host chiral electromagnetic waves. The magnetization that results in a momentum space separation of a pair of opposite chirality Weyl nodes is also responsible for the non-zero gyrotropy parameter in the system. It is then shown that a chiral electromagnetic wave can propagate in a region of space where the gyrotropy parameter changes sign. Such waves are analogs of quantum Hall edge states for photons.
\end{abstract}
\maketitle

\section{Introduction}
The Weyl semimetal is a new topological phase of matter recently proposed theoretically, \cite{bib:Wan, bib:Yang, bib:Burkov, bib:Xu}.
The band structure of the Weyl semimetal consists of points in momentum space at which the valence and conduction bands touch; for a review see \cite{bib:Volovik-Review}. Weyl points always appear in pairs of opposite chiralities, which is required by the fermion-doubling theorem, \cite{bib:NN}, and are separated in momentum space if the time reversal or inversion symmetries are broken.

One of the unique properties of Weyl semimetals is the chiral anomaly \cite{bib:Adler, bib:Bell}. It results in the non - conservation of the number of particles of the given chirality in the presence of the electromagnetic (EM) field. Another unique property of the Weyl semimetal is the semi - quantized anomalous Hall effect \cite{bib:Burkov}.

On the experimental side, the search for a Weyl semimetal is stimulated by the recent
experimental realization of a three dimensional Dirac semimetal \cite{bib:Exp1, bib:Exp2, bib:Exp3}. We wish to note that such a Dirac semimetal is not yet a Weyl semimetal, for this Dirac semimetal is chirality degenerate. Or in other words, the opposite chiralities are found at the same point in momentum and energy. The inversion or time-reversal symmetries have to be broken in the compounds for it to have the Weyl points.
Recently, the compound Cd$_{3}$As$_{2}$ \cite{bib:CdAs} was experimentally shown to have signs of negative magnetoresistance, which is believed to be due to chiral anomaly \cite{bib:Nielsen_NMR, bib:Spivak}. The crystal structure of such a material has a broken inversion symmetry resulting in Weyl points split in energy. 

\begin{figure}[top]
     \includegraphics[width=80mm]{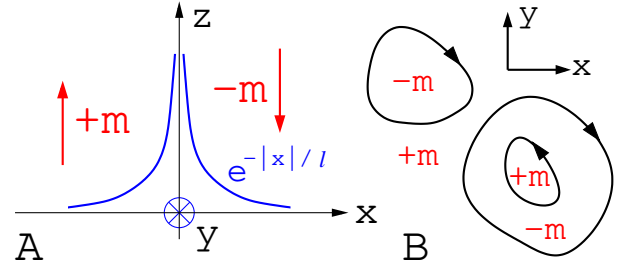}
     \caption{(Color online) A. Schematics of the domain wall separating two opposite magnetization $\pm {\vec m}$ (shown in red) at point $x=0$. The magnetization is parallel to the $z$-axis. The profile $\propto e^{-|x|/\ell}$ of localized electromagnetic wave at the domain wall and direction of propagation is shown in blue, where $\ell$ is the localization length of the wave. B. Schematics of the magnetization defined chirality of an electromagnetic wave. Black curves are the domain walls separating regions with different magnetization. Arrows are the directions of the propagation of the electromagnetic wave.
     }
\label{pic1}
\end{figure}
In the present paper we address an interesting question of propagation of an EM wave at the interface of Weyl semimetals.
For that, we adopt a particular model of a Weyl semimetal with broken time-reversal (TR) symmetry, which was introduced in Ref. \cite{bib:Burkov}. In this model the TR symmetry is broken by
the randomly distributed magnetic impurities, but which are assumed to be ferromagnetically ordered and result in the uniform exchange field for electrons.
Due to finite magnetization, Weyl points are split in momentum space, acquiring protection from perturbations that can open up a band gap. TR breaking also gives rise to the anomalous Hall effect proportional to the value of the splitting of the Weyl points \cite{bib:Burkov}. As a result, a Weyl semimetal is an optically gyrotropic media with gyrotropy parameter proportional to the sign of magnetization. Hence the Faraday rotation of the EM wave polarization is expected, \cite{bib:Jackiw, bib:Optics-Review}. 

We note that magnetic domain walls might appear naturally in the Weyl semimetal with broken TR symmetry or, for example, can be created with a help of a ferromagnetic material placed in proximity. The domain wall acts as an effective interface between right-left gyrotropic regions of the Weyl semimetal.

It is well known that the right-left gyrotropic crystal interface can localize chiral EM waves \cite{bib:Optics-Review, bib:Raikh}. Importantly, such chiral surface EM waves have attracted considerable theoretical \cite{bib:Haldane1, bib:Haldane2} and experimental \cite{bib:Wang, bib:Hafezi} attention as analogs of the quantum Hall state for photons. 

Here we propose a Weyl semimetal as another material to host such interesting chiral EM edge states. 
We show that in the region where the magnetization flips its directions (magnetic domain wall), there exists a chiral EM wave localized at the domain wall and propagating along it, with the direction of propagation determined by the sign of the gyrotropy parameter; see Fig. \ref{pic1} (a). Even though the system under consideration is a metal and the decay of the EM wave is expected, we show that there is a region of frequencies when the propagation is not damped.

\section{The model of Weyl semimetal}
We consider a simplest theoretical realization of a Weyl semimetal, based on a topological insulator - ferromagnetic insulator multilayer heterostructure grown along the $z$-axis \cite{bib:Burkov}. The model is well studied and we refer to Ref. \cite{bib:Burkov} for details. In this section we are going to briefly review the model with an emphasis on the effect of TR symmetry breaking.

Following \cite{bib:Burkov}, we introduce the exchange field $m(x)$ that acts on electrons, which physically could be due to magnetic impurities ordered ferromagnetically along the growth direction. To model the domain wall, we assume the magnetization to change the sign at the plane $x=0$, such as $m(x) = -|m| \mathrm{sign}(x)$.
We will first understand the properties of the Weyl semimetal with a homogeneous magnetization, namely $m(x) = m$, and will come back to the domain wall afterwards.

The momentum space Hamiltonian of the Weyl semimetal is given by:
\begin{eqnarray} \label{Ham2}
\mathcal{H}_{\pm}(\mathbf{ p}) =v_F\hat{z}\times \bsigma\cdot\mathbf{ p}+m\sigma^z + \hat{\Delta}(p_z)-\mu,~
\end{eqnarray}
where $v_F$ is the Fermi velocity, $\hat{z}$ is the growth direction of the heterostructure, $\sigma$ and $\tau$ are the real spin and the pseudospin describing the top and bottom surfaces of the layers, $m$ is the exchange field, $\hat{\Delta}(p_z) = \Delta_S\tau^x+  \Delta_D (
\tau^{+}e^{ip_zz} + h.c.)$ describes the motion in the growth direction, $\tau^{+} =( \tau^x\pm i \tau^y)/2$, and $\mu$ is the chemical potential.
Performing unitary transformation, 
\begin{equation}
U = \frac{1+\tau^z}{2} + \frac{1-\tau^z}{2} \sigma^z
\end{equation}
one obtains the Hamiltonian of the multilayer in the form:
\begin{eqnarray} \label{Ham2}
\mathcal{H}_{\pm}(\mathbf{ p}) =v_F\hat{z}\times \bsigma\cdot\mathbf{ p}+(m \pm\Delta(p_z))\sigma^z-\mu,~
\end{eqnarray}
where $\Delta(p_z)=[\Delta_S^2+\Delta_D^2+2\Delta_S\Delta_D\cos(p_zd)]^{1/2}$, and $d$ is the period of the heterostructure.

Weyl points can exist in the subband described
by $\mathcal{H}_{\pm}(\mathbf{ p})$ for the $\mathrm{sign}(m)=\mp 1$, respectively. At $m=0$ there is a single four-fold degenerate Weyl point provided $\Delta_S=\Delta_D$. We will not refer to this special limit here.

As long as the exchange energy satisfies the inequality $|\Delta_S-\Delta_D|<|m|<\Delta_S+\Delta_D$, the separation between two Weyl points in momentum space is given by:
\begin{equation}\label{Q}
2Q=\frac{2}{d}\textrm{arccos}\left[\frac{\Delta_S^2+\Delta_D^2-m^2}{2\Delta_S\Delta_D} \right]~.
\end{equation}
In what follows, we assume that the chemical potential is not far from the
Weyl points, $||m| - |\Delta_S-\Delta_D|| > |\mu|$, allowing us to neglect the contribution from the $\mathcal{H}_{\pm}(\mathbf{ p})$ band for $\mathrm{sign}(m)=\pm 1$, respectively.
Finally, one arrives at the Hamiltonian describing low energy excitations near the Weyl points,
\begin{eqnarray} \label{Ham3}
\mathcal{H}(\mathbf{ p}) =v_F\hat{z}\times \bsigma\cdot \mathbf{ p}+(|m| -\Delta(p_z))\mathrm{sign}(m)\sigma^z-\mu.~~
\end{eqnarray}
From here we observe that the sign change of magnetization flips the chirality of electrons. It is convenient to write the Green function of electrons in the Weyl semimetal $G(\epsilon_n,\mathbf{ p})=[i\epsilon_n-\mathcal{H}(\mathbf{ p})]^{-1}$ in the form:
\begin{equation}\label{Green}
G(\epsilon_n,\mathbf{ p})=\frac{1}{2}\sum_{s=\pm 1}\frac{1+s\bsigma\cdot \mathbf{ n}(\mathbf{ p})}{i\epsilon_n +\mu -sE(\mathbf{ p})},
\end{equation}
where the summation is performed over two subbands with opposite chirality, $\epsilon_n =\pi T(2n+1)$ is the fermionic Matsubara frequency, $T$ is the temperature, $E(\mathbf{ p})= \sqrt{v_F^2p_{\bot}^2+(|m|-\Delta(p_z))^2}$ is the band dispersion of the Weyl semimetal, and
$
\mathbf{ n}(\mathbf{ p})=\left\{v_Fp_y,-v_Fp_x,(|m|-\Delta(p_z))\mathrm{sign}(m)\right\}/E(\mathbf{ p})
$ is the unit vector locked to the direction of the momentum. With the help of the Green function we are now going to study the macroscopic properties of the system.

\section{ Dielectric function of the Weyl semimetal}
Here we derive the tensor of the dielectric function, which will allow us to understand the properties of the propagation of the EM wave in the media. The dielectric function is expressed through the optical conductivity: 
\begin{equation}
\sigma_{ab}(\omega)= \frac{i}{\omega}\lim_{\mathbf{ q}\rightarrow 0}[\Pi_{ab}(\omega,\mathbf{ q}) -\Pi_{ab}(0,\mathbf{ q})],
\end{equation}
where $a, b = (x,y,z)$, in the form: 
\begin{equation}
\epsilon_{ab}(\omega)= \delta_{ab}+\frac{i\sigma_{ab}(\omega)}{\varepsilon_0\omega},
\end{equation} 
and $\varepsilon_0$ is the permittivity of free space. The current-current correlation function reads:
\begin{eqnarray}\label{Pi}\nonumber
&&\Pi_{ab}(\omega,\mathbf{ q})=e^2T\sum_{n}\mathrm{Tr}\int \frac{d^3p}{(2\pi)^3}
G(\epsilon_{n}+\omega_k,\mathbf{ p}+\mathbf{ q})\\
&\times&[\partial_{p_{a}}\mathcal{H}(\mathbf{p})\vert_{\mathbf{p} + \mathbf{q}}]G(\epsilon_n,\mathbf{ p})[\partial_{p_{b}}\mathcal{H}(\mathbf{ p})]\vert_{i\omega_k\rightarrow \omega+i\delta}~,
\end{eqnarray}
where $\mathrm{Tr}$ is taken over the pseudo spin degrees of freedom, $\omega_k=2\pi kT$ is the bosonic external frequency, and $e$ is the elementary charge.
For the long wave-length EM field we neglect the momentum dependence of the dielectric function. 

Without losing the generality, we consider the chemical potential in the electron band $\mu \geqslant 0$ and take $\omega>0$. Since the chemical potential is set close to the Weyl nodes, we linearize the band dispersion $E(\mathbf{ p})$ in the $z$-direction in momentum space. At zero temperature, $T=0$, the tensor of the dielectric function of the Weyl semimetal has the following form:
\begin{equation}\label{dielectric_tensor}
\bepsilon(\omega)=\begin{pmatrix}
\epsilon_x(\omega)& i\gamma(\omega)&0\\
-i\gamma(\omega)& \epsilon_x(\omega)&0\\
0&0&\epsilon_z(\omega)\\
\end{pmatrix}.
\end{equation}
The diagonal components of the dielectric function $\bepsilon(\omega)$ are given by:
\begin{eqnarray}\label{diag}\nonumber
\epsilon_{a}(\omega)=1+\frac{\alpha c_a}{3\pi}\bigg[\ln\bigg|\frac{4\Lambda^2}{4\mu^2-\omega^2}\bigg| -\frac{4\mu^2}{\omega^2}
+i\pi \Theta(\omega-2\mu) \bigg],\\
\end{eqnarray}
where $\alpha=\frac{e^2}{4\pi\varepsilon_0 \tilde{v}_F}$ is the material dependent fine-structure constant, and $\Lambda$ is the energy cut-off satisfying $\Lambda \gg \omega$ and $\Lambda \gg \mu$.
Coefficients $c_x=1$ and $c_z=\tilde{v}_F^2/v_F^2$ take into account the anisotropy of the Fermi velocity in the multilayer. 

The imaginary term in expression (\ref{diag}) describes the interband contribution to the optical conductivity.
This contribution exists if the frequency of the EM wave is larger than twice the Fermi energy: $\omega>2\mu$. Importantly, we find the interval of frequencies in which the dielectric function is real and positive, $\mathrm{Im }\epsilon_x(\omega) =0$ and $\mathrm{Re}\epsilon_x(\omega) >0$:
\begin{equation}\label{region}
2\mu>\omega>\sqrt{\frac{\alpha}{3\pi}}\frac{2\mu}{[1+\frac{2\alpha}{3\pi} \ln|\Lambda/\mu| ]^{1/2}}
\end{equation}
This means that the EM wave can propagate in the Weyl semimetal without decay in the interval of frequencies (\ref{region}). In what follows, we will consider that the frequency of the EM wave is within this interval.

The gyrotropy parameter $\gamma(\omega)$ is related to the off-diagonal component of the dielectric tensor as $\epsilon_{xy}(\omega)=-\epsilon_{yx}(\omega)=i\gamma(\omega)$ and is proportional to the anomalous Hall conductivity of the Weyl semimetal with broken time reversal symmetry:
\begin{eqnarray}\label{gamma}
\gamma(\omega)&=&\gamma_0(\omega) \mathrm{sign}(m),
\end{eqnarray}
where $\gamma_0(\omega) =\frac{2\alpha}{\pi}\frac{\tilde{v}_FQ}{\omega}$. We note that $\gamma(\omega)$ is a real value provided $\omega+2\tilde{v}_FQ < 2\Lambda$, inversely proportional to the frequency of the EM wave, and proportional to the splitting between Weyl points in momentum space (\ref{Q}) and to the sign of the magnetization. We would like to emphasize here that the anomalous Hall conductivity in the Weyl semimetal and as a result the gyrotropy parameter are determined by the distance between the Weyl nodes.

\section{Chiral electromagnetic surface states}
Let us now study the propagation of the EM wave in the Weyl semimetal with broken TR symmetry. 
We will adopt the theoretical model for the EM waves at the interface of optical isomers studied by Zhukov and Raikh \cite{bib:Raikh}. 
The wave equation with the dielectric tensor (\ref{dielectric_tensor}) derived in the previous section is given by:
\begin{equation}\label{EM}
\bnabla\times\bnabla\times\mathbf{ E}(\omega, \mathbf{ r})=\frac{\omega^2}{c^2}\bepsilon(\omega)\mathbf{ E}(\omega, \mathbf{ r})~.
\end{equation}
We consider the polarization and the direction of the propagation of the electric field to be perpendicular to the $z$-axis and take into account condition (\ref{region}).
The dispersion of the electric field:
\begin{equation}
\mathbf{ E}(x,y) =
E\bigg(1,\frac{c^2q_y^2-\omega^2\epsilon_x(\omega)}{c^2q_x
q_y+i\omega^2 \gamma(\omega)},0\bigg)e^{ixq_x+iyq_y},
\end{equation}
where $E$ is the amplitude, in the bulk of Weyl semimetal is given by:
\begin{eqnarray}\label{qmax}
q^2_{x}+q^2_{y}=\frac{\omega^2}{c^2}\epsilon_x(\omega)\left[1-\frac{\gamma^2(\omega)}{\epsilon_x^2(\omega)}\right]~.
\end{eqnarray}
The gyrotropy parameter is typically smaller than $\epsilon_{x}(\omega)$, and thus the EM wave can propagate in the bulk of the semimetal.

Let us now assume that there is a domain wall separating regions with opposite magnetization. For example the domain wall is at the plane $x=0$, and the magnetization as a function of coordinates is $m(x) = -|m| \mathrm{sign}(x)$.
We have shown in expression (\ref{gamma}) that the gyrotropy parameter is directly proportional to the sign of magnetization. Hence the gyrotropy parameter also flips the sign with the magnetization: 
\begin{equation}
\gamma(\omega)\rightarrow \gamma(\omega,x)=-\gamma_0(\omega) \mathrm{sign}(x).
\end{equation}
One solves the resulting wave equation with $\gamma(\omega) \rightarrow \gamma(\omega, x)$ for the electric field propagating in the $x-y$ plane, normal to the plane of the domain wall, taking into account condition (\ref{region}). Assuming a homogeneous solution along the $y$-axis, 
\begin{equation}
\mathbf{ E}(x,y)= (E_x(x),E_y(x),0)e^{iqy}, 
\end{equation}
and expressing $E_x(x)$ through $E_y(x)$ results in the following equation:
\begin{eqnarray}\label{EM2}
-\partial_x^2E_y(x) -\frac{2q\gamma_0(\omega) }{\epsilon_x(\omega)}\delta(x)E_y(x)=\mathcal{E}(\omega)E_y(x)~,
\end{eqnarray}
where the term on the right-hand side of Eq. (\ref{EM2}) reads:
$\mathcal{E}(\omega)=- q^2+ \omega^2[\epsilon_x^2(\omega)-\gamma_0^2(\omega)]/c^2\epsilon_x(\omega)$.
We note that the delta-function potential is either attractive or repulsive depending on the sign of the wave-vector $q$.
Solving Eq. (\ref{EM2}), one obtains a bound-state solution only for $q>0$ with an energy $\mathcal{E}(\omega) = -(q\gamma_0(\omega)/\epsilon_x(\omega))^2~$, while the spectrum of the bound state satisfies equation $q=\omega\sqrt{\epsilon_x(\omega)}/c$.
The components of the electric field are given by:
\begin{subequations}
\begin{align}\label{Answer}
E_y(x,y) &= E_y(0)e^{iqy-|x|/\ell}~,\\
E_x(x,y) &= iE_y(x,y)\frac{\epsilon_x^2(\omega)+\gamma_0^2(\omega)}{2\epsilon_x(\omega)\gamma_0(\omega)}\mathrm{sign}(x)~.
\end{align}\label{Answer}
\end{subequations}
The electric field is localized at the domain wall with the decay length given by:
\begin{equation}
\ell = \frac{\pi c}{2\alpha\tilde{v}_F }\frac{\sqrt{\epsilon_x(\omega)}}{Q}.
\end{equation}
The localization length is inversely proportional to the splitting of Weyl nodes in momentum space. Importantly, the EM wave can propagate at the domain wall without decay provided condition \ref{region} is satisfied, i.e., the diagonal component of the dielectric tensor, $\epsilon_x(\omega)$, is real and positive.

The obtained bound state solution exists only for wave-vectors $q>0$. If we have chosen $m(x) = \vert m\vert {\text sign}(x)$ for the domain wall configuration, then the bound state would have existed only for $q<0$. Hence the EM wave given by Eqs. (\ref{Answer}a), (\ref{Answer}b) is chiral (propagating in one direction). Another feature is that due to the nature of the gyrotropy parameter, given by Eq. (\ref{dielectric_tensor}), we expect the propagation of such chiral EM wave only in the plane perpendicular to the magnetization, i.e., in the $(x-y)$ plane (see Fig. \ref{pic1} (b) for schematics).

We note that the wave-vector of the EM field at the domain wall is smaller than the maximal allowed wave-vector in the bulk of the semimetal defined in (\ref{qmax}). Thus, the amplitude of the EM wave at the domain wall exponentially decays into the bulk \cite{bib:Raikh}. The assumption $\omega \gg v_F q $ that allowed us to neglect the momentum dependence of the dielectric tensor (\ref{dielectric_tensor}), is fulfilled provided $\sqrt{\epsilon_x(\omega)}v_F/c \ll 1$. Finally, we emphasize that apart from the localized EM wave (\ref{Answer}), Eq. (\ref{EM}) provides a solution for the EM wave propagating in the bulk. On the contrary, the EM waves discussed in Refs. \cite{bib:Haldane1, *bib:Haldane2} do not coexist with the bulk waves as there is a gap in the spectrum for bulk waves.

\section{Conclusions}
Let us briefly discuss the experimental observability of the proposed effect. We take energy cut-off $\Lambda\sim 1$ $\mathrm{eV}$, chemical potential $\mu\sim 0.1$ $\mathrm{eV}$, and the wave-vector
$Q\sim 1$ $\mathrm{nm}^{-1}$ for the multilayer model \cite{bib:Burkov}. Using $\tilde{v}_F \sim 10^8$ $\mathrm{cm}/\mathrm{c}$ gives $\alpha \sim 2$ and $\epsilon_x\sim 1$ at frequencies $\omega \sim 2\pi \cdot16 $ $\mathrm{THz}$. We obtain the localization length to be of the order of $\ell \sim 1$ $\mu\mathrm{m}$.

To conclude, we studied the properties of the EM wave at the vicinity of the magnetic domain wall in the Weyl semimetal. We showed that the chiral EM wave is localized at the domain wall and propagates along it without decay, which is the analog of the quantum Hall edge states for photons. 

\section{Acknowledgements}

We thank A. A. Burkov and A. Yu. Zyuzin for reading the manuscript and for valuable questions. AAZ acknowledges support from the Swiss NF, NCCR QSIT, and RFFI under Grant No. 12-02-00300-A. VAZ acknowledges support from the National Science Foundation via Grant No. NSF DMR-1308972.

\bibliography{Weyl-PRB-Ref2}

\end{document}